\documentclass[aps,prxquantum,twocolumn,superscriptaddress,nofootinbib]{revtex4-2}

\usepackage{graphicx}
\usepackage{amsmath,amssymb}
\usepackage{xcolor}
\definecolor{linkblue}{rgb}{0.0,0.2,0.6}
\usepackage[colorlinks=true, linkcolor=linkblue, citecolor=linkblue, urlcolor=linkblue]{hyperref}

\begin{document}

\title{Modular Neural Decoder for Surface Code Memory\\ and Logic on Superconducting Processors}

\author{Moritz Lange}
\email{moritz.lange@physics.gu.se}
\affiliation{Department of Physics, University of Gothenburg, Gothenburg, Sweden}
\author{Ole Fjeldså}
\affiliation{Department of Physics, Chalmers University of Technology, Gothenburg, Sweden}
\author{Gustaf Jonasson Johansson}
\affiliation{Department of Physics, Chalmers University of Technology, Gothenburg, Sweden}
\author{Mats Granath}
\email{mats.granath@physics.gu.se}
\affiliation{Department of Physics, University of Gothenburg, Gothenburg, Sweden}

\date{\today}

\begin{abstract}

Neural decoders for quantum error correction have achieved high accuracy on surface code memory experiments, but practical fault-tolerant computation requires decoders that transfer across devices, noise conditions, and logical operations. We introduce a modular graph-recurrent decoder designed for this setting. Built from reusable local syndrome encoders, it is pretrained in simulation and fine-tuned to hardware using a common pipeline, with only lightweight spatial aggregation adapted between experiments. The decoder performs competitively on conventional memory benchmarks, outperforming minimum-weight perfect matching (MWPM) at distances 3 and 5 on an IBM Nighthawk-generation processor and closely approaching AlphaQubit~\cite{bausch2024alphaqubit} on Google's below-threshold hardware data~\cite{google2024threshold}. We then test the framework in two settings beyond memory decoding on the same IBM device. First, in an $X$-type stability experiment, we continuously tune the effective measurement error rate using a variable pre-readout rotation and train a single decoder jointly across the resulting noise range. It matches or outperforms MWPM across the sweep without requiring a separately calibrated decoder for each setting. Second, we decode a $ZZ$-merge lattice-surgery experiment between two distance-3 surface-code patches, improving the decoded correlation between the merge outcome and the final $ZZ$ parity relative to MWPM on real hardware. Together, these experiments demonstrate neural decoding that is portable not only across hardware, but also across qualitatively different surface code observables and logical operations, using models trainable on single-GPU academic resources.

\end{abstract}

\maketitle

\section{Introduction}
\label{sec:intro}

Quantum error correction (QEC) encodes logical information redundantly in many physical qubits and extracts error information through repeated stabilizer measurements~\cite{dennis2002topological,fowler2012surface, gottesman1997stabilizercodesquantumerror}. The surface code is a leading candidate for superconducting processors, and recent experiments have demonstrated the key milestone of operation below threshold: logical error rates that decrease as the code distance $d$ grows~\cite{google2024threshold}. Realizing this suppression in practice, however, depends, among other things, on the classical decoder, the algorithm that infers logical corrections from the measured syndrome record: decoding Google's distance-7 memory experiment~\cite{google2024threshold} with plain minimum-weight perfect matching on the basic pairwise $p_{ij}$-calibrated prior gives an error-suppression factor of only $\Lambda \approx 1.56$, well below the $\Lambda \approx 2.14$ AlphaQubit reaches on the same data~\cite{google2024threshold, bausch2024alphaqubit}. Closing that gap is not the only requirement, however: whichever decoder is used must also stay fast enough to keep pace with the rate at which syndromes are measured, around 1~$\mu$s for superconducting qubits~\cite{Terhal_2015, google2024threshold}.

Machine learning-based decoders have developed from solving toy decoding problems \cite{torlai2017neural,meinerz2022scalable,gu2026scalable, Chamberland_2018, Baireuther_2019, Krastanov_2017, wang2023transformerqecquantumerrorcorrection, Andreasson_2019, Fitzek_2020, Sweke_2020} to learning the noise model directly from experimental data \cite{Varbanov_2025,lange2025gnn, zhang2026parallel}. The state of the art is AlphaQubit, a recurrent transformer that outperformed any classical decoder on Google's surface code memory experiments~\cite{bausch2024alphaqubit} and has since been scaled to larger codes with real-time throughput~\cite{senior2025alphaqubit2}. Parallel work has focused on low-latency architectures and hardware integration \cite{Overwater_2022, chamberland2023techniques,gicev2023scalable,zhang2026parallel,yon2025memristive,cicero2026lowlatencygnnaccelerator,yang2026realtime}. The crucial question we address in this work is whether a neural decoder can perform well on real data \emph{portably}: across hardware platforms, noise profiles, and different amounts of available data, and across logical operations beyond the standard memory experiment, all without re-engineering the decoder for each setting.

In response to this challenge we introduce a modular neural decoder. The architecture, which extends our earlier graph-neural-network decoder~\cite{lange2025gnn} and the hybrid graph-recurrent model of Ref.~\cite{fjeldsa2025thesis}, follows a renormalization-group-inspired design~\cite{duclos2010fast}: a single learned encoder, processing a $3\times3\times2$ space-time block of detection events, is reused at every scale, its per-block embeddings coarse-grained by a small convolutional layer and integrated over time by a recurrent network. Each new code distance, or adjacent surface code patches during lattice surgery, adds only a small convolutional aggregation module, so the parameter count is essentially independent of $d$; nearly every model in this paper contains a few million parameters and was trained on a single GPU.

\begin{figure*}
\includegraphics[width=\textwidth]{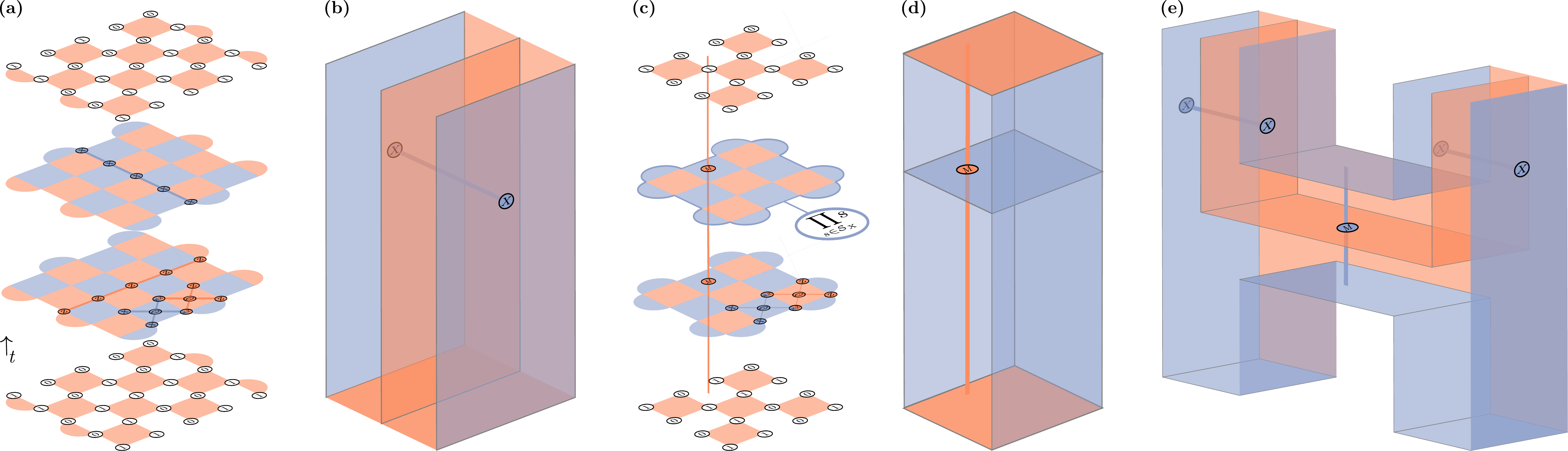}
\caption{Surface code patches (a,c) and their spacetime pictures (b,d,e)~\cite{zhang2026hitchhiker}. Orange (blue) faces are $Z$($X$)-type boundaries and logical surfaces; the latter represent the tracked logical observable. (a,b) Memory-$Z$ experiment: the $Z$-basis initialization and measurement form $Z$-type temporal boundaries, and the tracked $Z_L$ surface spans the full duration of the experiment (top and front $Z$-faces omitted for clarity). (c,d) Stability experiment: the patch has $X$-type boundary stabilizers on all four spatial sides, and the tracked surface is instead an $X$-stabilizer product lying flat in space, at a single round (front $X$-face omitted for clarity). (e) $ZZ$-merge: two patches, each initialized in $|{+}\rangle$ and finally measured in $Z$ ($|{+}{+}\rangle\to ZZ$, mirroring the real hardware demonstration of Sec.~\ref{sec:merge}), share a row of stabilizers for a merge window, corresponding to the measurement of the joint operator $Z_1Z_2$ (orange surface spanning the merge window in space). Because each patch is initialized with an $X$-type boundary, neither $Z_1$, $Z_2$, nor the merge outcome $m$ is fixed individually; only the combination $Z_1\oplus Z_2\oplus m$ forms a closed observable (Table~\ref{tab:merge_constraints}), corresponding to the Bell-pair correlation decoded in Sec.~\ref{sec:merge} (top and front $Z$-faces omitted for clarity).}
\label{fig:background}
\end{figure*}

The remainder of the paper is organized as follows. Sec.~\ref{sec:background} introduces the surface code memory, stability, and lattice surgery experiments. Sec.~\ref{sec:architecture} describes the modular decoder architecture and the three-stage fine-tuning pipeline. Sec.~\ref{sec:results} presents our results, organized by experiment type: memory experiments (Sec.~\ref{sec:memory}) on IBM and Google hardware, stability experiments including tunable measurement errors (Sec.~\ref{sec:stability}), and lattice surgery merge decoding (Sec.~\ref{sec:merge}). We discuss the results in Sec.~\ref{sec:discussion} and conclude in Sec.~\ref{sec:conclusion}.

We summarize our main results below:
\begin{enumerate}
\item \emph{Memory on IBM hardware.} Fine-tuned on data from IBM hardware, the decoder beats MWPM at $d=3$ and at $d=5$ (Sec.~\ref{sec:ibm}).
\item \emph{Google hardware.} On the public data of the below-threshold experiment~\cite{google2024threshold}, the decoder reaches logical error rates on the $Z$ basis experiments that closely match the industrial-scale AlphaQubit decoder (Sec.~\ref{sec:google}).
\item \emph{Beyond memory experiments.} The same decoding architecture decodes the $X$-type stability experiment~\cite{gidney2022stability}, the structural proxy for the merge step of lattice surgery~\cite{horsman2012lattice}, on real IBM hardware, beating MWPM at every duration tested (Sec.~\ref{sec:temporal}).
\item \emph{Tunable measurement errors.} Using the stability circuit as a testbed, we sweep a variable $R_X(\alpha)$ rotation on every measure qubit before readout, similar to the protocol used for memory experiments in Ref.~\cite{google2024threshold}. A single decoder fine-tuned jointly across all 17 $(t,\alpha)$ settings beats MWPM (Sec.~\ref{sec:tunable}).
\item \emph{Lattice surgery merge decoding.} Finally, the same architecture decodes a full lattice surgery $ZZ$ merge between two distance-3 surface code patches on real hardware, successfully decoding the resulting Bell-pair experiment and outperforming MWPM (Sec.~\ref{sec:merge}).
\end{enumerate}

\section{Surface code background: memory, stability, and lattice surgery experiments}
\label{sec:background}

The surface code encodes one logical qubit in a two-dimensional lattice of \emph{data} and \emph{measure} qubits~\cite{dennis2002topological,fowler2012surface}. Data qubits sit on the lattice vertices; measure qubits, one per plaquette, repeatedly measure four-qubit (or, at the boundary, two-qubit) $X$- or $Z$-type stabilizers arranged in a checkerboard, so that every bulk data qubit participates in two $X$-type and two $Z$-type stabilizers. A stabilizer outcome that differs from its value in the previous round is a \emph{detection event}, and it is from the pattern of detection events accumulated over many stabilizer measurement rounds that a decoder infers whether a logical observable has been flipped. Which logical observable(s) can be defined in a given experiment depends on the boundary conditions, as well as initialization and measurement bases, as we describe next.

\subsection{Memory versus stability experiments}

\begin{table*}[t]
\caption{The $2^4-1=15$ nontrivial combined constraints generated by the four independent $ZZ$-merge checks $C_{Z_1}$, $C_{Z_2}$, $C_{ZZ}$, and $C_{X_1X_2}$, written out as explicit XOR combinations of the initialization/measurement outcomes, $z(x)_i^{\rm in/out}$, and the merge measurement result $m$. The last five columns mark which $2^k-1$ constraints apply with: both qubits initialized and measured in $Z$ ($k=3$); qubit 1 (2) initialized in $Z$ ($X$) and both measured in $Z$ ($k=2$); qubit 1 (2) initialized in $Z$ ($X$) and measured in $X$ ($Z$) ($k=1$); both qubits initialized and measured in $X$ ($k=1$); and both qubits initialized in $X$ and measured in $Z$ ($k=1$).}
\label{tab:merge_constraints}
\begin{ruledtabular}
\begin{tabular}{lcccccc}
Logical check & Constraint & $|00\rangle\!\to\!ZZ$ & $|0{+}\rangle\!\to\!ZZ$ & $|0{+}\rangle\!\to\!XZ$ & $|{+}{+}\rangle\!\to\!XX$& $|{+}{+}\rangle\!\to\!ZZ$ \\
\hline
$C_{Z_1}$ & $z_1^{\rm in}\oplus z_1^{\rm out}$ & \checkmark & \checkmark & & &\\
$C_{Z_2}$ & $z_2^{\rm in}\oplus z_2^{\rm out}$ & \checkmark & & & &\\
$C_{ZZ}$ & $z_1^{\rm in}\oplus z_2^{\rm in}\oplus m$ & \checkmark & & & &\\
$C_{X_1X_2}$ & $\Xi\equiv x_1^{\rm in}\oplus x_1^{\rm out}\oplus x_2^{\rm in}\oplus x_2^{\rm out}$ & & & & \checkmark &\\
$C_{Z_1}\oplus C_{Z_2}$ & $z_1^{\rm in}\oplus z_1^{\rm out}\oplus z_2^{\rm in}\oplus z_2^{\rm out}$ & \checkmark & & & &\\
$C_{Z_1}\oplus C_{ZZ}$ & $z_1^{\rm out}\oplus z_2^{\rm in}\oplus m$ & \checkmark & & & &\\
$C_{Z_1}\oplus C_{X_1X_2}$ & $z_1^{\rm in}\oplus z_1^{\rm out}\oplus \Xi$ & & & & &\\
$C_{Z_2}\oplus C_{ZZ}$ & $z_1^{\rm in}\oplus z_2^{\rm out}\oplus m$ & \checkmark & \checkmark & \checkmark & &\\
$C_{Z_2}\oplus C_{X_1X_2}$ & $z_2^{\rm in}\oplus z_2^{\rm out}\oplus \Xi$ & & & & &\\
$C_{ZZ}\oplus C_{X_1X_2}$ & $z_1^{\rm in}\oplus z_2^{\rm in}\oplus m\oplus \Xi$ & & & & &\\
$C_{Z_1}\oplus C_{Z_2}\oplus C_{ZZ}$ & $z_1^{\rm out}\oplus z_2^{\rm out}\oplus m$ & \checkmark & \checkmark & & & \checkmark\\
$C_{Z_1}\oplus C_{Z_2}\oplus C_{X_1X_2}$ & $z_1^{\rm in}\oplus z_1^{\rm out}\oplus z_2^{\rm in}\oplus z_2^{\rm out}\oplus \Xi$ & & & & &\\
$C_{Z_1}\oplus C_{ZZ}\oplus C_{X_1X_2}$ & $z_1^{\rm out}\oplus z_2^{\rm in}\oplus m\oplus \Xi$ & & & & &\\
$C_{Z_2}\oplus C_{ZZ}\oplus C_{X_1X_2}$ & $z_1^{\rm in}\oplus z_2^{\rm out}\oplus m\oplus \Xi$ & & & & &\\
$C_{Z_1}\oplus C_{Z_2}\oplus C_{ZZ}\oplus C_{X_1X_2}$ & $z_1^{\rm out}\oplus z_2^{\rm out}\oplus m\oplus \Xi$ & & & & &\\
\end{tabular}
\end{ruledtabular}
\end{table*}

A conventional surface code patch has two pairs of boundaries of opposite type, e.g.\ $Z$-type top and bottom and $X$-type left and right, so that a logical $Z_L$ ($X_L$) operator can terminate on the two $Z$-type ($X$-type) boundaries (Fig.~\ref{fig:background}(a)). The \emph{memory experiment} initializes and measures the data qubits in one basis, say $Z$, and asks whether the corresponding logical observable $Z_L$ is preserved through multiple rounds of stabilizer measurements with data, measurement, and gate errors. In the spacetime picture the patch is a tube whose temporal faces (initialization and final measurement) are $Z$-type, matching the two $Z$-type spatial boundaries. In practical terms, data qubit errors update the physical Pauli frame, corresponding to where in the full Hilbert space of the code the logical qubit is stored. Modulo these changes, which are irrelevant for the logical information, a logical bit flip corresponds to a chain of $X$ data errors which pierces the tracked $Z_L$ surface, as pictured in Fig.~\ref{fig:background}(b). The task of the decoder is to make the binary prediction of whether a logical bit flip has occurred or not based on the set of recorded detection events.     

A \emph{stability experiment}~\cite{gidney2022stability} inverts this picture. Choosing a patch with the \emph{same} boundary type all the way around its perimeter, e.g.\ $X$-type (Fig.~\ref{fig:background}(c)), means every data qubit participates in exactly two $X$-type stabilizers, so the product of all $X$-type stabilizers is a constant known ahead of time, $\prod_{s \in \mathcal{S}_X} s = +1$. This fixed product is itself the true value of the observable: rather than tracking whether a logical operator survives through time, the experiment asks whether the $+1$ value can be correctly inferred from the detection record, without providing the relevant syndrome information that defines the measured value of the observable. Initializing and measuring the data qubits in the \emph{opposite} basis from the perimeter, $Z$ in this example, makes verifying that global product nontrivial: a chain of measurement errors on the $X$-stabilizers spanning the duration of the experiment can flip it without ever producing a detection event. Correspondingly, in the spacetime picture (Fig.~\ref{fig:background}(d)) the tracked observable surface lies flat in space, rather than spanning the full time-like extent of the tube as in (b). 

A memory experiment tests whether an observable survives being carried \emph{forward in time}, and a stability experiment tests whether a logical observable can be moved \emph{across space}. Together, these experiments capture the mechanisms underlying logical operations between multiple surface code patches with \emph{lattice surgery}.

\subsection{Lattice surgery experiments and success criteria}
\label{sec:merge_targets}

Lattice surgery implements joint logical Pauli measurements by temporarily merging surface code patches and measuring stabilizers across their shared boundary. These measurements provide the building blocks for logical gates such as CNOT~\cite{horsman2012lattice,fowler2019lowoverheadquantumcomputation, Litinski_2018, Litinski_2019, bodeker2026lattice}. Lattice surgery has been demonstrated with trapped ions~\cite{erhard2021entangling,steiner2026genuine}, neutral atoms~\cite{bluvstein2025architectural}, and superconducting qubits~\cite{besedin2026lattice,lacroix2025color,wang2026surgery,lin2026logical}. Assessing their success requires checking both the reported measurement outcome and the preservation of compatible logical information. For a joint $ZZ$ measurement between two logical qubits, we can choose the following four independent generators of logical checks: consistency of the $ZZ$ measurement outcome with the input $Z$-basis parity $C_{ZZ}=z_1^{\rm in}\oplus z_2^{\rm in}\oplus m$, preservation of $Z_1$ and $Z_2$ individually, $C_{Z_i}=z_i^{\rm in}\oplus z_i^{\rm out}$, and preservation of $X_1X_2$, $C_{X_1X_2}=x_1^{\rm in}\oplus x_1^{\rm out}\oplus x_2^{\rm in}\oplus x_2^{\rm out}$. Here $z(x)_i^{\rm in/out}\in \{0,1\}$ are the initialization and measurement outcomes on individual logical qubits in the two bases and $m\in \{0,1\}$ the merge measurement result, given by the product of measured $Z$-stabilizers in the merge region. 

The four generators yield $2^4-1=15$ nontrivial combinations. In the absence of errors these should all be 0. However, a particular experiment can test only those combinations compatible with its initialization and readout bases.  For product-state Pauli preparation and single-qubit Pauli readout, every remaining input term must have a known preparation eigenvalue, and every remaining output term must be accessible from the measurement record. Nevertheless, a combination can remain testable even when its constituent generators are not, because their incompatible boundary terms can cancel. As exemplified in Table~\ref{tab:merge_constraints}, for various combinations of initialization and readout settings, this leaves $2^k-1$ nontrivial testable combinations corresponding to $k<4$ independent logical checks. Of particular relevance is the hardware experiment of Sec.~\ref{sec:merge}, illustrated in Fig.~\ref{fig:background}(e): both patches are initialized in $|+\rangle_L$ and finally measured in $Z$. Neither preservation of the individual $Z_i$ values nor agreement of the merge outcome with the initial $Z_1Z_2$ parity can be tested separately. Their combination nevertheless gives the deterministic relation $C_{Z_1}\oplus C_{Z_2}\oplus C_{ZZ}=z_1^{\rm out}\oplus z_2^{\rm out}\oplus m=0$, where the unavailable initial values cancel. This constraint tests agreement between the merge outcome and the final $ZZ$ parity of the resulting Bell pair.

In the following section we describe the decoder architecture and training, before moving on to results on memory, stability, and lattice surgery experiments. 

\section{Decoder architecture and training pipeline}
\label{sec:architecture}

\begin{figure*}[t]
\includegraphics[width=\textwidth]{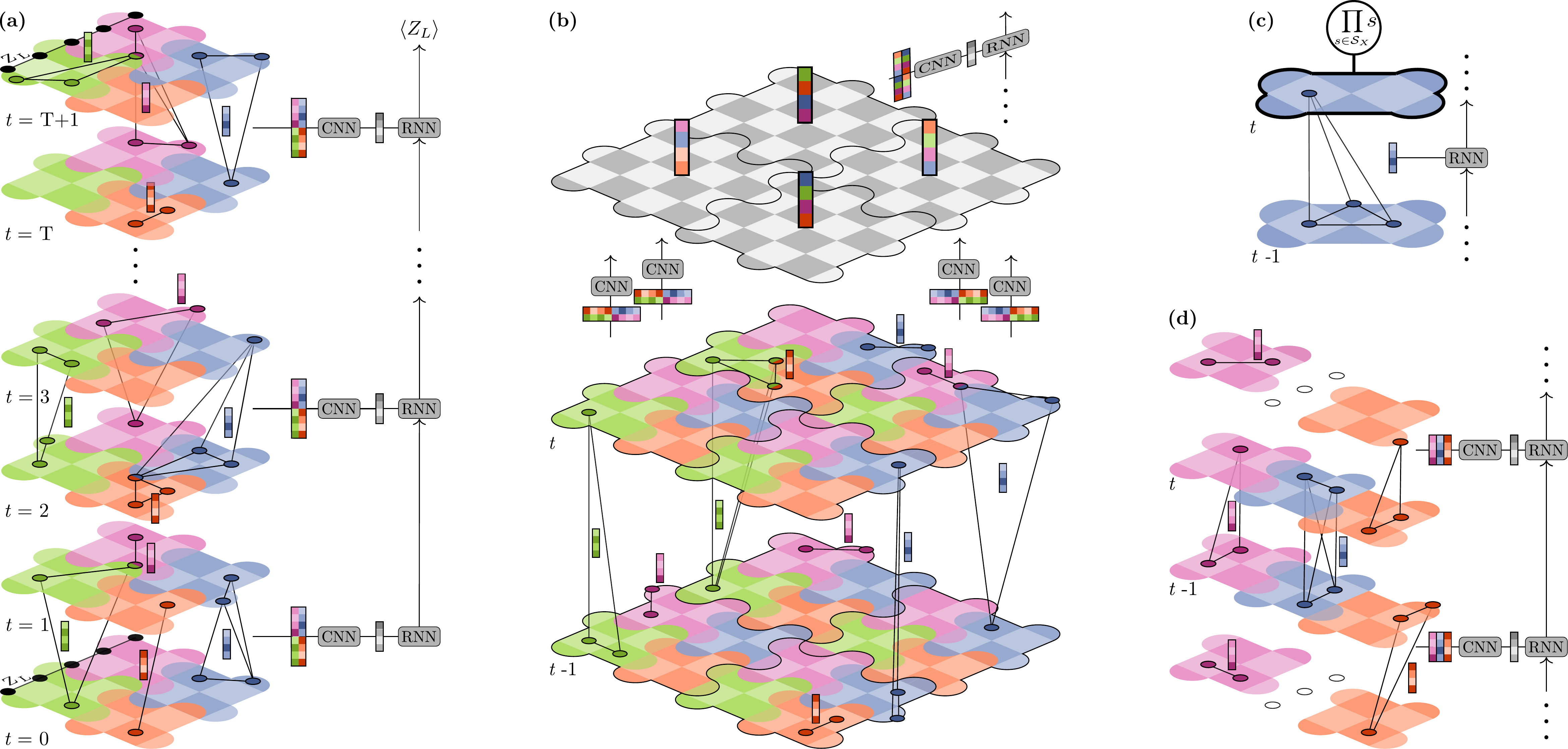}
\caption{Modular neural decoder for (a) a $d=5$ memory experiment, (b) a $d=9$ memory experiment, (c) a stability experiment, and (d) a $ZZ$ merge. Colors distinguish the overlapping $d=3$ tiles. Within each $\Delta t = 2$ temporal window, the detection events of each tile form a graph, which a shared GNN encodes into a single vector (colored bars). (a) A $2\times 2$ convolution (CNN) aggregates the four tile encodings into one embedding per window; an RNN (GRU) integrates these over time, and a linear head predicts the logical-flip probability (black nodes: support of $Z_L$). (b) At $d=9$ the construction is nested: four $d=5$ modules as in (a), each covering four $d=3$ tiles, produce embeddings that a second $2\times 2$ CNN aggregates on the $d=9$ lattice (gray) before the RNN. (c) The stability patch is decoded by a single base decoder without aggregation, predicting the product $\prod_{s\in\mathcal{S}_X} s$ of all $X$-type stabilizers. (d) The merged strip is split into three $d=3$ tiles aggregated by a $1\times 3$ CNN; before and after the merge only the two separate surface code patches are present.}

\label{fig:architecture}
\end{figure*}

\subsection{Base decoder}

The base decoder is a standalone $d=3$ graph-recurrent decoder (GNN-RNN). On its own, it operates on \emph{syndrome graphs}~\cite{lange2025gnn}: each detection event within a sliding temporal window of size $\Delta t = 2$ becomes a node annotated with its space-time coordinates. The nodes are connected all-to-all, with edge weights decaying as the inverse square of the supremum-norm distance between their space-time coordinates, resembling the decay in probability of chains of i.i.d.\ error events with increasing length. A message-passing GNN \emph{encodes} each window's graph into a fixed-length vector; windows without detection events contribute a learned ``empty'' embedding. The sequence of window embeddings is processed by a multi-layer gated recurrent unit (GRU), and a linear head maps the output to the probability that the logical observable flipped (which observable depends on the experiment, Sec.~\ref{sec:background}). Because the GNN acts on individual windows and the GRU carries all temporal context, the decoder generalizes to experiment durations far beyond those seen in training; we train at $t = 50$ rounds and evaluate up to $t = 1000$ (Appendix~\ref{app:memory_sim}).

\subsection{Modular decoder}
\label{sec:modular}

In the bulk of the surface code lattice, every stabilizer has the same neighborhood structure up to translation. This permits a renormalization-style decomposition~\cite{duclos2010fast}: a distance-$(2k{-}1)$ code splits into four overlapping distance-$k$ tiles that share boundary stabilizers along their seams, each of which closely resembles (though, due to boundary effects from sitting inside the larger lattice, is not exactly identical to) the structure of a standalone distance-$k$ code. Each tile is processed independently, and the results are combined into an effective description one level up. For $d=5$, we achieve this by running the \emph{same} GNN encoder of the pretrained $d=3$ decoder on every tile and coarse-graining the four tile encodings with a $2\times 2$ convolution, whose output feeds the GRU and prediction head (Fig.~\ref{fig:architecture}(a)). This recursion extends to $d = 9$ by nesting: four $d=5$ modules feed a $d=9$ module, and so on (Fig.~\ref{fig:architecture}(b)). $d=7$ does not fit the $2k{-}1$ recursion; instead, nine $d=3$ tiles cover the code on a $3\times 3$ grid, sharing seam stabilizers along tile boundaries the same way, aggregated by a $3\times 3$ convolution, see Appendix~\ref{app:architecture} for details.

A standalone $d=3$ decoder has $\approx 1.6$ million parameters. Composing it into a larger decoder reuses this entire base decoder, GNN encoder and GRU alike (with its weights continuing to update during composite training), and adds only one small convolutional aggregation module per additional level ($\approx 0.3$ million parameters per $2\times 2$ aggregation used at $d=5,9$; $\approx 0.66$ million for the $3\times 3$ aggregation used at $d=7$). The resulting parameter count is therefore nearly independent of $d$: $\approx 1.9$ million at $d=5$, $\approx 2.3$ million at $d=7$ and $d=9$, small enough that every model in this paper, up to $d=9$ on simulated and $d=7$ on real data, was trained in at most a few days on a single GPU. The one exception is a wider variant of the $d=7$ decoder, with the GRU hidden size raised from 256 to 1024 ($\approx 30$ million parameters), which we use in Sec.~\ref{sec:google} to test how accuracy at larger $d$ depends on the capacity of the recurrent state.

Nothing in the base decoder is specific to the memory experiment. Its input, the sequence of per-window syndrome graphs, is built identically from the detection events of any surface code circuit, and the choice of experiment in Sec.~\ref{sec:background} enters only through (i) the binary label that the linear head is trained to predict and (ii) the spatial layout of the surface code patches that the modular decoder tiles. For the memory experiment (Fig.~\ref{fig:background}(a,b)) the label is the flip of the tracked logical operator $Z_L$ between initialization and final measurement. For the stability experiment (Fig.~\ref{fig:background}(c,d)) the patch differs from a $d=3$ memory patch: with $X$-type boundaries on all four sides it has a different shape and a different number and arrangement of $X$- and $Z$-type stabilizers (compare the stability and memory patches in Fig.~\ref{fig:layout}). Because the GNN acts on graphs of detection events labeled by their space-time coordinates rather than on a fixed stabilizer grid, the base decoder architecture nevertheless applies without modification; it is simply trained on stability circuits; the label indicates whether the measured product of all $X$-type stabilizers deviates from its ideal value $+1$ (Fig.~\ref{fig:architecture}(c)). For the $ZZ$ merge (Fig.~\ref{fig:background}(e)) the label is the combined check $C_{Z_1}\oplus C_{Z_2}\oplus C_{ZZ}=z_1^{\rm out}\oplus z_2^{\rm out}\oplus m$ of Sec.~\ref{sec:merge_targets}. Other initialization and readout choices would change only which checks of Table~\ref{tab:merge_constraints} are available as labels, and a decoder with several output heads would predict all $k$ independent checks of a given experiment simultaneously. The merged region is a strip of width $2d$ plus the bridge, rather than a square patch, so we adapt the aggregation layer compared to the memory experiments: the strip is split along its long axis into three overlapping $d=3$ tiles (patch~1, the bridge region, patch~2), each embedded by a $d=3$ GNN encoder, and the $2\times 2$ (or $3\times 3$) convolution is replaced by a $1\times 3$ convolution across this row before the GRU and prediction head. Before and after the merge, the bridge contributes learned empty embeddings, just like a window without detection events in a memory experiment (Fig.~\ref{fig:architecture}(d)). 

\subsection{Three-stage fine-tuning pipeline}
\label{sec:pipeline}

All hardware results use the following training pipeline~\cite{bausch2024alphaqubit}:

\emph{Stage 1 --- synthetic pretraining.} The decoder is trained on Stim-simulated circuits~\cite{gidney2021stim}, either circuit-level depolarizing noise ($p \in \{0.001, \ldots, 0.005\}$) or a superconducting-inspired SI1000 noise model \cite{Gidney_2021} with error rates scaled to match the target device's observed detection-event density.

\emph{Stage 2 --- calibrated error model.} From the device's real detection events we fit correlation probabilities $p_{i_1 i_2 \ldots i_k}$ up to order $k=4$ and build a detector error model \cite{Spitz_2018, Remm_2025, chen2022calibrated}. The decoder is fine-tuned on synthetic samples generated from this model \cite{bausch2024alphaqubit}.

\emph{Stage 3 --- real data.} Finally the decoder is fine-tuned directly on the recorded hardware detection events, split 80/10/10 into train, validation, and a held-out test set from which all reported numbers derive. Appendix~\ref{app:val_from_start} shows how much of this validation accuracy is reached during each stage. All syndrome extraction cycle counts are mixed within every optimizer step, which preserves the temporal generalization of the base model.

Logical error rates $p_L$ are reported as failures over total shots at a given duration $t$; fitting $p_L(t) = \tfrac{1}{2}\left[1 - (1-2\varepsilon_d)^{t}\right]$ across all measured $t$ then yields the per-round rate $\varepsilon_d$ at code distance $d$. MWPM baselines use PyMatching~\cite{higgott2023sparse} on the same Stage-2 $p_{ij}$ detector error models, so the matching decoder benefits from device-informed edge weights.

\subsection{Compute cost and code availability}

Every model in this paper was trained on a single NVIDIA GH200 GPU, from fine-tuning runs of a few GPU-hours up to the largest models trained on simulated circuits, which take on the order of a few GPU-days. Code, training scripts, and all models are available at \url{https://github.com/LangeMoritz/modular-gnn-rnn-decoder}.

\section{Results}
\label{sec:results}

\subsection{Memory experiments}
\label{sec:memory}

\subsubsection{IBM hardware}
\label{sec:ibm}

\begin{figure}
\includegraphics[width=\columnwidth]{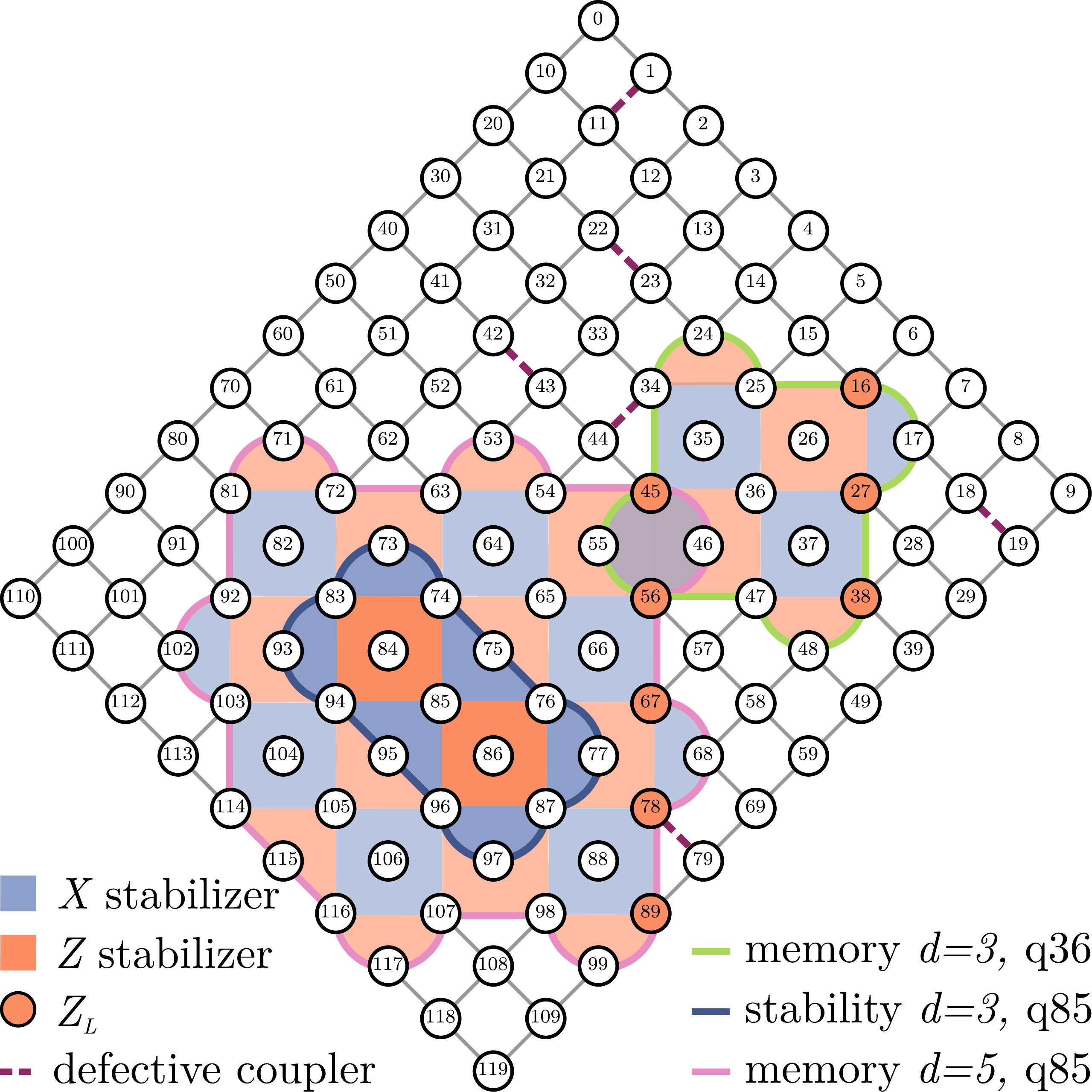}
\caption{Placement of the surface code patches on the \texttt{ibm\_berlin} qubit grid, numbered by physical index. White circles: data qubits; orange/blue tiles: $Z$/$X$-type measure qubits. Dashed purple lines mark couplers found dead or degraded during our hardware runs. Shaded regions: the $d=5$ memory patch (q85) (pink boundary, its inner-core $d=3$ memory patch omitted for clarity), the independent $d=3$ memory patch (q36) (green boundary), and the $d=3$ stability patch (blue boundary). Qubits on which we define and measure the logical $Z$ are marked in orange.}
\label{fig:layout}
\end{figure}

Previous QEC experiments on IBM hardware used earlier device generations~\cite{wootton2018repetition,chen2022calibrated,sundaresan2023matching,gupta2024magic,hetenyi2024creating,lee2026scalable}; surface code memory and stability experiments~\cite{drouet2026error,granath2026avkodning} as well as lattice surgery~\cite{lilja2026lattice} were recently run on the square-lattice Nighthawk r1 generation. We ran three surface code memory experiments on the IBM Nighthawk r1 processor \texttt{ibm\_berlin}: two independent $d=3$ regions and one $d=5$ region. Fig.~\ref{fig:layout} shows the physical placement of three of the four hardware patches (each patch labeled by the physical index of its center data qubit under the chip's row-major numbering): the $d=5$ memory patch (q85), the independent $d=3$ memory patch (q36), and the $d=3$ stability patch (q85) used for the stability experiment of Fig.~\ref{fig:stability}. The fourth, the $d=3$ memory patch that is the $d=5$ patch's inner core (also centered on q85), is omitted from the figure for clarity. Two constraints limited where a patch could sit. First, on this device generation, qubits with odd row+column parity suffer vastly degraded mid-circuit readout, so every measure qubit in a viable patch must have even row+column. Second, dead and degraded couplers on this device (dashed purple lines in Fig.~\ref{fig:layout}) ruled out most candidate regions. Therefore the region that remains large enough to host a $5\times5$ patch with even-parity measure qubits free of those defects is pinned against the edge of the chip. We resolved the corner constraint using the defect-adaptation approach of Ref.~\cite{leroux2025snakes} (see Ref.~\cite{drouet2026error} for a comparison of defect-handling strategies on the same processor generation): we removed the corner data qubit and its neighboring boundary measure qubit, leaving the adjacent boundary measure qubit q115 at weight 3 instead of weight 4. This reduces the code distance for the logical $X$ observable to four, while leaving the complementary logical observable ($Z$, the basis used for the memory-experiment results of Table~\ref{tab:ibm}) at the full distance of five.

Mid-circuit reset is slow on this device, so we use a scheme where measure qubits are measured but never reset between rounds. We recover the standard per-round detection event by comparing outcomes two rounds apart \cite{geher2025reset}. With $10^5$ shots at each duration $t \in \{5, 10, 20, 30, 50\}$, we applied the pipeline of Sec.~\ref{sec:pipeline}: Stage 1 pretrains on SI1000 circuits with noise scaled to each patch's detection-event density, Stages 2 and 3 fine-tune on the patch's calibrated detector error model and real detection events, respectively.

\begin{figure}
\includegraphics[width=\columnwidth]{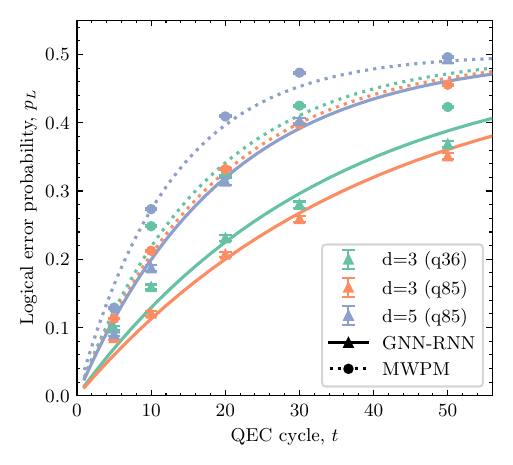}
\caption{Logical error rate versus experiment duration for surface code memory experiments on IBM hardware (\texttt{ibm\_berlin}), GNN-RNN (solid, triangles) versus MWPM (dotted, circles) on the identical held-out hardware shots, for three patches: $d=3$ (q36), $d=3$ (q85), and $d=5$ (q85).}
\label{fig:ibm_d3}
\end{figure}

\begin{table*}
\caption{Fitted error per round $\varepsilon_d$ on IBM \texttt{ibm\_berlin} memory experiments (held-out test shots), GNN-RNN versus MWPM decoding of the same shots, for all three patches (labeled by center-qubit index; see Sec.~\ref{sec:ibm}).}
\label{tab:ibm}
\begin{ruledtabular}
\begin{tabular}{lcccccc}
 & \multicolumn{2}{c}{$d=3$ (q36)} & \multicolumn{2}{c}{$d=3$ (q85)} & \multicolumn{2}{c}{$d=5$ (q85)} \\
 & GNN-RNN & MWPM & GNN-RNN & MWPM & GNN-RNN & MWPM \\
\hline
$\varepsilon_d$ & 1.47\% & 2.79\% & 1.26\% & 2.60\% & 2.48\% & 3.79\% \\
\end{tabular}
\end{ruledtabular}
\end{table*}

Table~\ref{tab:ibm} and Fig.~\ref{fig:ibm_d3} show the results for all three patches. At $d=3$ the GNN-RNN roughly halves the logical error per round relative to MWPM on both independent regions ($1.47\%$ vs.\ $2.79\%$ at q36; $1.26\%$ vs.\ $2.60\%$ at q85). At $d=5$ the modular decoder again beats MWPM at every duration except $t=50$, where both approach the random-guessing ceiling. Growing the q85 core from $d=3$ to $d=5$ nearly doubles the detection-event density ($16\%\to28\%$), and the fitted error per round nearly doubles for the GNN-RNN ($1.26\%\to2.48\%$) and grows by about half for MWPM ($2.60\%\to3.79\%$), the opposite of the below-threshold suppression seen on Google hardware.

\subsubsection{Google hardware}
\label{sec:google}

\begin{figure*}
\includegraphics[width=\textwidth]{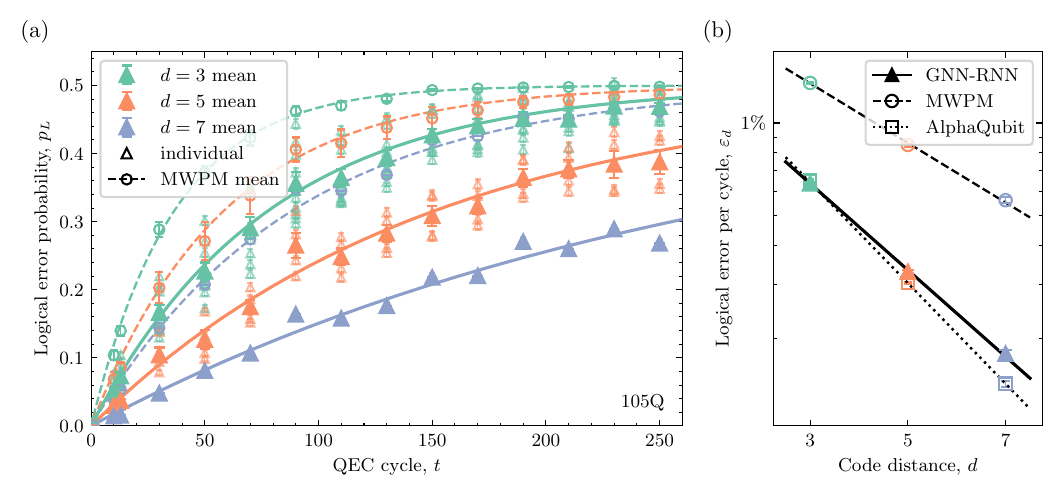}
\caption{(a) Logical error probability $p_L$ versus QEC cycles $t$ and (b) logical error per round $\varepsilon$ versus code distance $d$ on the Google surface code memory data ($Z$ basis), after the full Stage-1/2/3 fine-tuning pipeline, against an MWPM baseline (patch mean, decoded on the same calibrated $p_{ij}$ detector error models) and Google's own AlphaQubit numbers.}
\label{fig:google}
\end{figure*}

We apply the same pipeline as for the IBM data to the public data of Google's below-threshold experiment~\cite{google2024threshold}: $Z$-basis memory experiments on a 105-qubit processor comprising nine $d=3$ patches, four $d=5$ patches, and one $d=7$ patch. The dataset consists of experiments of duration $t = 1, 10, 13, 30, 50, 70, 90, 110, 130, 150, 170, 190, 210, 230, 250$, each with 50{,}000 shots per patch and basis, which we split 80/10/10 into training, validation, and test sets. The $d=3$ patches use the flat base decoder; $d=5$ and $d=7$ use the modular architecture with the shared $d=3$ GNN front end ($2\times 2$ and $3\times 3$ aggregation, respectively). At $d=7$ we train both the default model and the wider variant of Sec.~\ref{sec:modular}.

Fig.~\ref{fig:google} summarizes the outcome: $\varepsilon_3 \approx 0.638\%$, $\varepsilon_5 \approx 0.330\%$, $\varepsilon_7 \approx 0.179\%$ (wide model), averaged over patches at each distance. Error suppression continues through the largest available distance, with $\varepsilon_3/\varepsilon_5 \approx 1.94$ and $\varepsilon_5 /\varepsilon_7 \approx 1.84$ ($\Lambda \approx 1.9$ from an exponential fit). MWPM~\cite{higgott2023sparse} on the same per-$t$ calibrated $p_{ij}$ detector error models gives $\varepsilon_3 \approx 1.35\%$, $\varepsilon_5 \approx 0.849\%$, $\varepsilon_7 \approx 0.562\%$ and $\Lambda \approx 1.56$.

AlphaQubit, trained on the same data with the same three-stage protocol, reaches $\varepsilon_3 = 0.616\%$, $\varepsilon_5 = 0.276\%$ and $\varepsilon_7 = 0.130\%$ in the $Z$ basis~\cite{google2024threshold}. Our decoder is within 4\% of it at $d=3$, but the gap grows to 20\% at $d=5$ and 38\% at $d=7$. A growing gap of this kind is what one would expect from a fixed-size recurrent state. In the modular decoder, each window's spatial information is compressed into one 256-dimensional vector before the GRU, so the temporal memory available per stabilizer falls as $1/d^2$. AlphaQubit instead keeps a separate recurrent state for each stabilizer. The $d=7$ capacity test supports this picture. With the default width, the decoder reaches $\varepsilon_7 \approx 0.215\%$, 65\% above AlphaQubit. Widening the GRU fourfold lowers this to $0.179\%$, so about 40\% of the gap closes from added recurrent capacity alone, with the spatial modules unchanged.

\subsection{Stability experiments}
\label{sec:stability}
\subsubsection{Temporal scaling}
\label{sec:temporal}

\begin{figure*}
\centering
\includegraphics[width=0.48\textwidth]{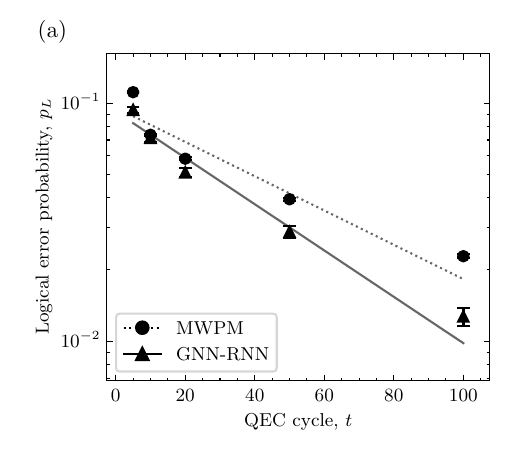}\hfill
\includegraphics[width=0.48\textwidth]{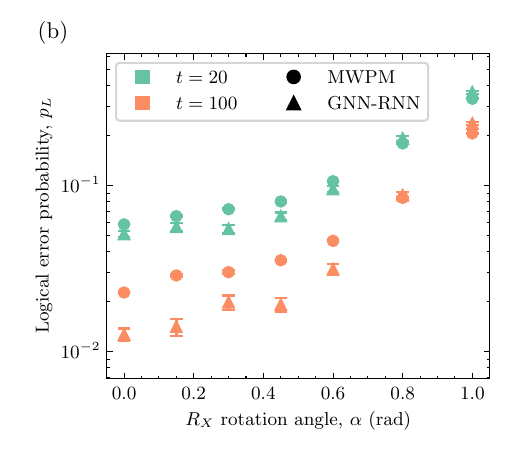}
\caption{$X$-type stability experiment on \texttt{ibm\_berlin}, GNN-RNN (triangles) versus MWPM (circles). (a) Logical error rate $p_L$ versus number of rounds $t$ at native noise ($\alpha = 0$). Lines are exponential fits, Eq.~\eqref{eq:lambda_t}, giving $\Lambda_t \approx 1.09$ (GNN-RNN) and $1.07$ (MWPM) per four rounds. (b) $p_L$ versus coherent $R_X(\alpha)$ rotation angle applied to all measure qubits, at $t = 20$ (teal) and $t = 100$ (orange), using a single GNN-RNN fine-tuned jointly across all $\alpha$.}
\label{fig:stability}
\end{figure*}

We ran the $d=3$ $X$-type stability circuit on \texttt{ibm\_berlin} ($10^5$ shots at each of $t \in \{5, 10, 20, 50, 100\}$) and fine-tuned the simulated stability model through the same Stage-2/Stage-3 pipeline as the memory experiments. Fig.~\ref{fig:stability}(a) shows the held-out test results: the GNN-RNN beats MWPM at every duration, with the relative advantage growing with $t$ (a 15\% error reduction at $t=5$, growing to 45\% at $t=100$).

Unlike a memory experiment, where $p_L$ grows with $t$ at fixed $d$, here $t$ plays the role $d$ plays for memory: a stability experiment's time-like code distance scales with the number of rounds, so $p_L$ decreases with $t$. Mirroring the $\Lambda = \varepsilon_d/\varepsilon_{d+2}$ convention of Fig.~\ref{fig:google}, a naive fit would take the exponent $(t-t_0)/2$. However, due to slow mid-circuit reset, as in the memory experiments above, we run a no-reset scheme: measure qubits are measured but never reset, and each per-round detection event is recovered by comparing outcomes two physical rounds apart~\cite{geher2025reset}. In this scheme a logical error only needs $t/2$ conspiring measurement errors to be undetectable, leading to the following phenomenological expression
\begin{equation}
p_L(t) = p_L(t_0)\,\Lambda_t^{-(t-t_0)/4}, \qquad t_0 = 5\,.
\label{eq:lambda_t}
\end{equation}
Fig.~\ref{fig:stability}(a) shows a fit to the data, giving $\Lambda_t \approx 1.09$ for the GNN-RNN versus $1.07$ for MWPM per four rounds.

\subsubsection{Tunable measurement errors}
\label{sec:tunable}

Finally, we introduce a new protocol for the stability circuit, inspired by the equivalent memory experiment \cite{google2024threshold}: a variable measure qubit rotation that lets us sweep the effective measurement error rate smoothly on hardware, rather than being limited to a device's fixed native noise. Inserting $R_X(\alpha)$ on every measure qubit immediately before its readout realizes an additional syndrome-measurement flip of probability $\sin^2(\alpha/2)$. We swept $\alpha \in \{0.15, \ldots, 1.0\}$ rad at $t = 20$ and $t = 100$ on \texttt{ibm\_berlin} ($5\times 10^4$ shots per point). The hardware reproduces the simulated phenomenology (Fig.~\ref{fig:stability_sim}): $p_L$ rises monotonically with $\alpha$, and the $t=100$ curve stays below the $t=20$ curve at every $\alpha$ (Fig.~\ref{fig:stability}(b)). Whereas MWPM requires its edge weights to be re-derived from a separately calibrated detector error model at every setting, the neural decoder can absorb all settings into a single training set. We therefore fine-tune one decoder jointly across all 17 $(t,\alpha)$ settings, with Stage 2 on the calibrated detector error models of every setting and Stage 3 on the real shots. Although this decoder is trained on a uniform mixture rather than specialized to each angle, it matches or beats the per-setting-calibrated MWPM at every duration and rotation angle tested.

\subsection{Lattice surgery experiments}
\label{sec:merge}

\begin{figure}
\includegraphics[width=\columnwidth]{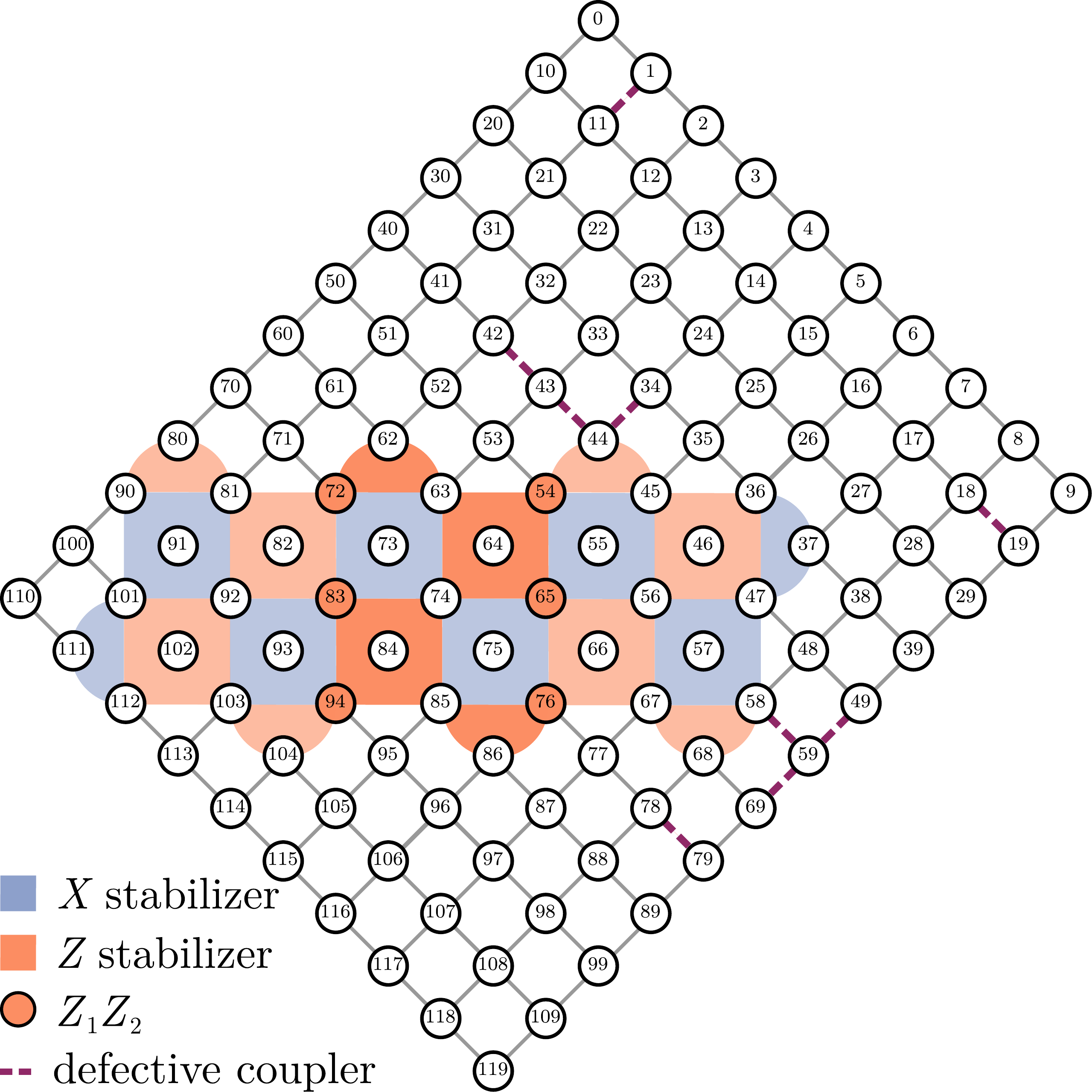}
\caption{Placement of the $ZZ$-merge experiment on the \texttt{ibm\_berlin} qubit grid, numbered by physical index. White circles: data qubits; blue/orange tiles: $X$/$Z$-type measure qubits. Dashed purple lines mark couplers found dead or degraded during our hardware runs. They differ from those in Fig.~\ref{fig:layout}: the merge and memory/stability experiments were run on different dates, and \texttt{ibm\_berlin}'s set of dead and degraded couplers drifts between recalibrations. Circled qubits form the representative $Z_1$ and $Z_2$ logical columns whose product gives the $Z_1Z_2$ merge observable.}
\label{fig:merge_layout}
\end{figure}

Sec.~\ref{sec:merge_targets} described how the logical correctness of a $ZZ$ merge can be checked in terms of four independent constraints, and combinations thereof, depending on the chosen initialization and measurement bases. Here we test the modular decoder on a $ZZ$ merge experiment over two $d=3$ surface code patches, which in the ideal case prepares a Bell pair. The two surface code patches are prepared in $|{+}{+}\rangle$, merged with a $ZZ$ measurement, and measured out in $Z$ (Fig.~\ref{fig:background}(e)). As shown in Table~\ref{tab:merge_constraints}, the relevant observable for this experiment is $z_1^{\rm out}\oplus z_2^{\rm out}\oplus m$ which depends only on the final measurements and the merge parity. In the ideal case it is 0, corresponding to the creation of a Bell pair $|{0}{0}\rangle+|{1}{1}\rangle$ ($|{0}{1}\rangle+|{1}{0}\rangle$) for $m=0$ ($m=1$). We decode this circuit with three overlapping GNN encoders, centered on the left surface code patch, the bridge region, and the right surface code patch, respectively, as described in Sec.~\ref{sec:modular}.

\begin{figure*}
\includegraphics[width=\textwidth]{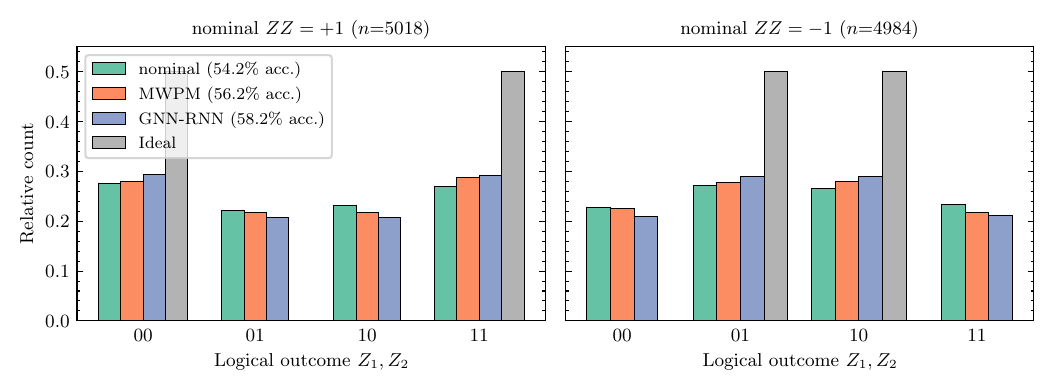}
\caption{$ZZ$-merge lattice surgery decoding, GNN-RNN versus MWPM. Bell-state histogram on real \texttt{ibm\_berlin} hardware (rounds 1--3--1, held-out test set, $n=10{,}002$), split by the merge measurement's nominal outcome; the joint $(Z_1, Z_2)$ outcome should ideally land only in the corresponding (00, 11) or (01, 10) outcomes of the $ZZ=\pm1$ panels shown. GNN-RNN correction reaches 58.2\% accuracy versus 56.2\% for MWPM (54.2\% uncorrected).}
\label{fig:merge_results}
\end{figure*}

We ran one round of individual surface-code stabilizer measurements, followed by three rounds on the merged patch, including the newly introduced bridge stabilizers, and a final round of individual measurements. The experiment was performed on \texttt{ibm\_berlin} with $10^5$ shots using the physical-qubit layout shown in Fig.~\ref{fig:merge_layout}. We then fine-tuned the simulated model using the same Stage-2/Stage-3 pipeline as for the memory and stability experiments. On a held-out test set ($n=10{,}002$, never used for training or checkpoint selection), the corrected outcome distribution moves toward the theoretical Bell-pair correlation (Fig.~\ref{fig:merge_results}): GNN-RNN correction reaches 58.2\% accuracy, ahead of MWPM (56.2\%) and the uncorrected outcomes (54.2\%). The corrections from both decoders stay far from the ideal histogram here, reflecting this device's high native noise rather than a decoder-specific limitation; the margin by which the GNN-RNN leads MWPM is nonetheless consistent with its advantage in the matched simulated sweep in Appendix~\ref{app:merge_sim}.

\section{Discussion}
\label{sec:discussion}

The combined results establish the versatility of our modular neural decoder: the same basic architecture, using a convolution over local GNN encoders together with a recurrent temporal unit, and the same three-stage training pipeline, produce MWPM-beating decoding on two hardware platforms and on three different logical-observable experiments. Neural decoders trained on experimental data have mostly been demonstrated on surface code memory experiments~\cite{Varbanov_2025,bausch2024alphaqubit,senior2025alphaqubit2,zhang2026parallel}, and only recently on logical operations such as lattice surgery~\cite{lacroix2025color,lin2026logical}, typically with a decoder trained for each experimental configuration. Here, the components themselves are reused: a single local syndrome encoder, pretrained on $d=3$ patches, is shared across memory, stability and merge experiments, and each new experiment changes only the training label and a small aggregation layer, while the training pipeline stays the same.

Neural decoders are currently being developed with different priorities: large models such as AlphaQubit~\cite{bausch2024alphaqubit,senior2025alphaqubit2} reach the highest accuracies on experimental data, while compact networks target the microsecond latencies of real-time decoding~\cite{Overwater_2022,gicev2023scalable,yon2025memristive,cicero2026lowlatencygnnaccelerator,yang2026realtime}, and it has been argued that near-term accuracy depends more on the amount of training data than on architectural complexity~\cite{yan2026rethink}. With a few million parameters and single-GPU training, our decoder is small enough that the full pipeline can be rerun for every patch, experiment type and noise setting, which is what made the stability sweep and the lattice-surgery experiment practical within an academic compute budget.

The Google results (Sec.~\ref{sec:google}) suggest that recurrent capacity becomes increasingly important at larger code distances. A natural next step is to give each tile its own recurrent state ahead of the spatial aggregation, so that capacity grows with the number of tiles while the modular structure and the reuse of the pretrained $d=3$ encoder are kept; related local, weight-shared designs have been explored in AlphaQubit~2~\cite{senior2025alphaqubit2} and in convolutional decoders~\cite{gu2026scalable}. Seed ensembling and larger pretraining sets offer further, complementary improvements.

Our Google results cover the $Z$ basis; training the $X$ basis is straightforward but not yet done. IBM results stop at $d=5$, set by the currently available patch quality and size: the same-qubit q85 comparison (Sec.~\ref{sec:ibm}) shows that the $d=5$ patch is noisier than the $d=3$ core it contains, so error suppression with distance on IBM hardware awaits better devices, not a different decoder. We report no latency or throughput benchmarks. The per-round cost of the architecture is constant in experiment duration, which is compatible with, but does not by itself demonstrate, real-time operation. Real-time neural decoding has meanwhile been demonstrated on TPUs~\cite{senior2025alphaqubit2,zhang2026parallel} and FPGAs~\cite{yang2026realtime}. For our GNN, we have previously shown sub-microsecond decoding per round on an FPGA for a fixed time window~\cite{cicero2026lowlatencygnnaccelerator}; extending this to streaming multi-round syndrome data is the natural next step.

\section{Conclusion}
\label{sec:conclusion}

We have presented a versatile modular neural decoder which has been applied to a range of surface code experiments, including memory, stability, and lattice surgery, as demonstrated through experiments on superconducting quantum hardware. Logical accuracies in all three instances match or outperform those of a hardware-noise-informed matching decoder. The modular structure of the decoder makes it readily adaptable to any near-term surface code experiment. The moderate size of the neural network implies that the required training compute is manageable on a single GPU and can facilitate portability to hardware systems dedicated to decoding streaming data. The decoder is purely data-driven and does not require the decoding problem to admit a matching formulation, making it applicable in principle to a broader class of quantum error-correcting codes. Future work should explore this potential beyond the surface code and extend the approach to the larger code sizes expected on next-generation quantum hardware.

\begin{acknowledgments}

M.L.\ and M.G.\ acknowledge support from the Knut and Alice Wallenberg Foundation through the Wallenberg Centre for Quantum Technology (WACQT). Computational resources were provided by the National Academic Infrastructure for Supercomputing in Sweden (NAISS), funded by the Swedish Research Council. Access to IBM hardware was provided by the WACQT Quantum Technology Testbed operated by Chalmers Next Labs, which is funded by the Knut and Alice Wallenberg Foundation. We thank Fang Zhang for helpful discussions on the lattice surgery merge experiments.
\end{acknowledgments}

\appendix

\section{Simulation}
\label{app:simulated}

\subsection{Memory}
\label{app:memory_sim}

\begin{figure}
\includegraphics[width=\columnwidth]{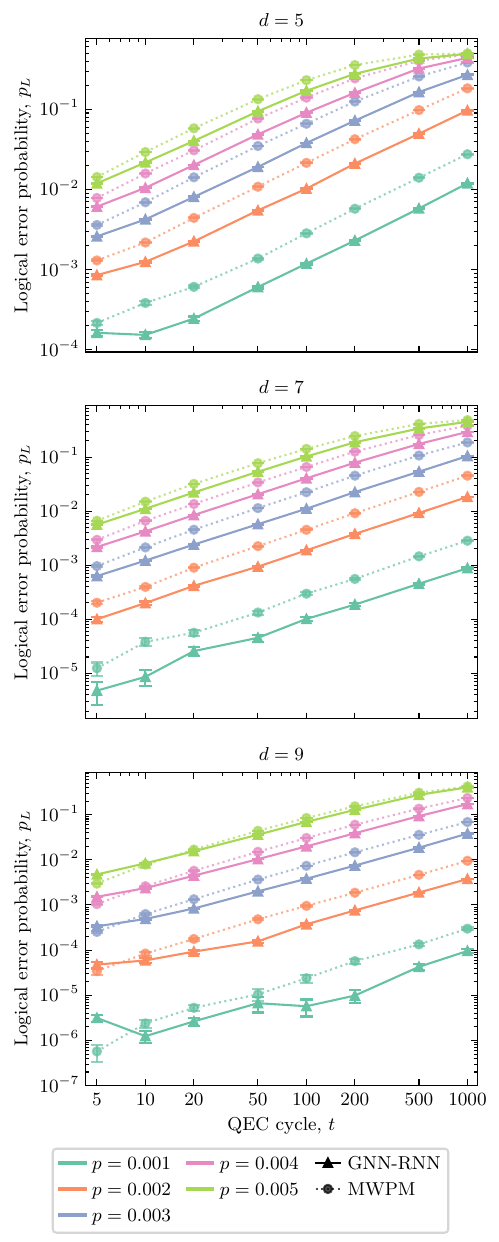}
\caption{Logical error rate versus number of rounds under circuit-level depolarizing noise for $d = 5, 7, 9$ and $p = 0.001$--$0.005$, GNN-RNN (triangles) versus MWPM (circles); lines are a guide to the eye. The modular decoder is trained at $t = 50$ rounds and evaluated out to $t = 1000$.}
\label{fig:simulated}
\end{figure}

We benchmark the modular decoder under circuit-level depolarizing noise, training each level on a mixture of physical error rates $p \in \{0.001, \ldots, 0.005\}$ at $t = 50$ rounds and testing at $t = 5$--$1000$ rounds (Fig.~\ref{fig:simulated}). At every distance tested, $d = 5, 7, 9$, the decoder beats MWPM across the full range of error rates and durations, with the advantage holding over several orders of magnitude in $p_L$. The important structural point is that going from $d=5$ to $d=9$ adds only one more copy of the same small CNN module already used at $d=5$: increasing accuracy at growing distance comes from the modular architecture without substantial model growth. The stability of the GNN-RNN advantage up to $t = 1000$, twenty times the training duration, confirms that the design generalizes temporally, a prerequisite for decoding the long syndrome streams of real algorithms.

\subsection{Stability}
\label{app:stability_sim}

\begin{figure}
\centering
\includegraphics[width=\columnwidth]{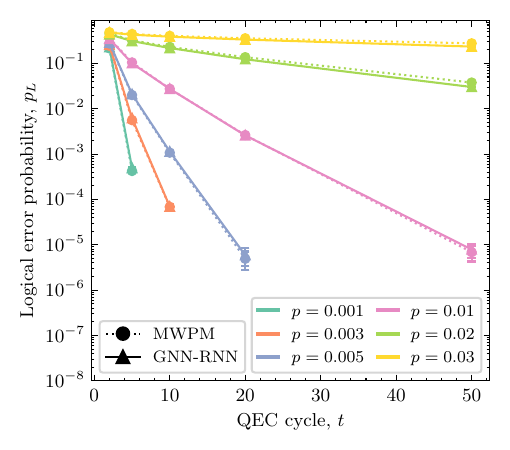}
\caption{Simulated $d=3$ $X$-type stability experiment under circuit-level depolarizing noise, GNN-RNN (triangles, solid) versus MWPM (circles, dotted), for a single decoder trained across all $p$.}
\label{fig:stability_sim}
\end{figure}

Trained across $p = 0.001$--$0.03$ under depolarizing noise, a $d=3$ stability decoder reproduces the expected qualitative behavior (Fig.~\ref{fig:stability_sim}): $p_L$ \emph{decreases} exponentially with the number of rounds $t$, the signature of the time-like code distance growing with $t$. At low physical error rates the GNN-RNN closely tracks MWPM, and it pulls ahead as $p$ increases, with the advantage widening at longer durations.

\subsection{Lattice surgery}
\label{app:merge_sim}

\begin{figure}
\includegraphics[width=\columnwidth]{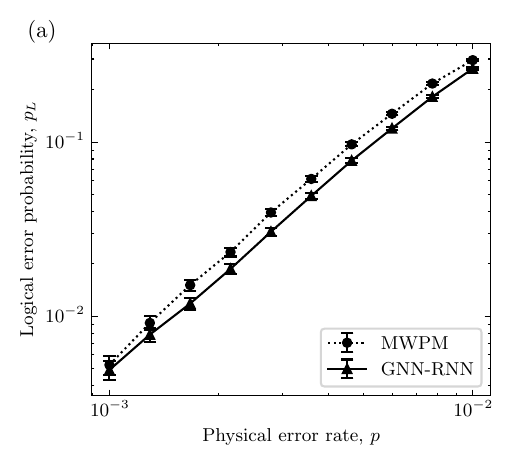}
\caption{$ZZ$-merge lattice surgery decoding, GNN-RNN versus MWPM. Simulated noise sweep under circuit-level depolarizing noise (rounds 3--3--3, gap-1 bridge), the modular decoder beating MWPM uniformly over $p \in [0.001, 0.01]$.}
\label{fig:merge_results_sim}
\end{figure}

Under circuit-level depolarizing noise, with a 3--3--3 (initialization-merge-final) round schedule, the modular decoder beats MWPM uniformly across the whole sweep (Fig.~\ref{fig:merge_results_sim}).

\section{Decoder architecture in detail}
\label{app:architecture}

This appendix gives the full forward pass of the architecture summarized in Sec.~\ref{sec:architecture} and Fig.~\ref{fig:architecture}. Unless noted otherwise, dimensions are the defaults used throughout the paper: node feature dimension $3$, GNN widths $3 \to 64 \to 256$, GRU hidden size $256$ with $L = 4$ layers, and level-aggregation width $H_{\rm agg} = 256$ ($H_{\rm agg} = 1024$ for the wide $d=7$ variant of Sec.~\ref{sec:google}).

\subsection{Syndrome graph}

Within a sliding temporal window of $\Delta t$ rounds ($\Delta t = 2$ throughout this paper), every detection event becomes a node $i$ with feature vector $\mathbf{x}_i^{(0)} = (x_i, y_i, \tau_i) \in \mathbb{Z}^3$, its spatial coordinates on the code and its local time index $\tau_i \in \{0, \ldots, \Delta t - 1\}$ within the window. Nodes are connected all-to-all (self-loops excluded) with a symmetric edge weight given by the inverse-square supremum-norm distance between their space-time coordinates $\mathbf{x}_i^{(0)}$:
\begin{equation}
w_{ij} = \big\| \mathbf{x}_i^{(0)} - \mathbf{x}_j^{(0)} \big\|_\infty^{-2}, \qquad i \neq j.
\label{eq:edge_weight}
\end{equation}
A window with no detection events produces no graph at all; it is represented by a learned placeholder in Eq.~\eqref{eq:empty}.

\subsection{GNN encoding}

Two graph convolutional layers~\cite{morris2019weisfeiler} map each node's feature vector through widths $3 \to 64 \to 256$. For layer $\ell \in \{0, 1\}$, with weight matrices $\Theta_1^{(\ell)}, \Theta_2^{(\ell)}$ and bias $b^{(\ell)}$,
\begin{equation}
\begin{split}
\mathbf{x}_i^{(\ell+1)} = \mathrm{ReLU}\Big( &\Theta_1^{(\ell)} \mathbf{x}_i^{(\ell)} \\
&{}+ \Theta_2^{(\ell)} \sum_{j \in \mathcal N(i)} w_{ij}\, \mathbf{x}_j^{(\ell)} + b^{(\ell)} \Big)
\end{split}
\label{eq:graphconv}
\end{equation}
with $w_{ij}$ from Eq.~\eqref{eq:edge_weight} and $\mathcal N(i)$ the full node set of the graph (excluding $i$). The window's encoding is the mean of the final-layer node features, or a learned vector $e_\varnothing \in \mathbb{R}^{256}$ if the window contains no detection events:
\begin{equation}
h_c =
\begin{cases}
\dfrac{1}{n_c} \displaystyle\sum_{i=1}^{n_c} x_i^{(2)}, & n_c > 0, \\[1.2em]
e_\varnothing, & n_c = 0,
\end{cases}
\label{eq:empty}
\end{equation}
where $c$ indexes the window and $n_c$ its node count. An experiment of $t$ rounds carries detectors at $t+1$ distinct time coordinates (one final set of perfect $Z$($X$) stabilizers constructed from the measurement of the data qubits in the $Z$($X$) basis); sliding the $\Delta t$-wide window one round at a time across them gives $T = (t+1) - \Delta t + 1 = t - \Delta t + 2$ windows. Stacking their embeddings gives $h_{1:T} \in \mathbb{R}^{T \times 256}$, the input to the recurrent stage.

\subsection{Recurrent integration (GRU)}

An $L$-layer gated recurrent unit with $256$-dimensional hidden state processes $h_{1:T}$. Writing $u_t^{(1)} = h_t$ and, for layer $\ell = 1, \ldots, L$, hidden state $s_t^{(\ell)} \in \mathbb{R}^{256}$ (with $u_t^{(\ell+1)} = s_t^{(\ell)}$ the input to the next layer), the standard GRU update at each layer is
\begin{align}
r_t &= \sigma\!\left(W_{ir} u_t + b_{ir} + W_{hr} s_{t-1} + b_{hr}\right), \\
z_t &= \sigma\!\left(W_{iz} u_t + b_{iz} + W_{hz} s_{t-1} + b_{hz}\right), \\
n_t &= \tanh\!\left(W_{in} u_t + b_{in} + r_t \odot (W_{hn} s_{t-1} + b_{hn})\right), \\
s_t &= (1 - z_t) \odot n_t + z_t \odot s_{t-1},
\label{eq:gru}
\end{align}
where $r_t, z_t, n_t$ are the reset gate, update gate, and candidate hidden state; $W_{i\bullet}, W_{h\bullet}$ and $b_{i\bullet}, b_{h\bullet}$ the layer's learned input-side and hidden-side weight matrices and biases (one pair per gate); $\sigma$ the logistic sigmoid; and $\odot$ the element-wise product. Layer superscripts on $r_t, z_t, n_t, s_t, u_t$ and all weights are omitted for readability ($s_t^{(\ell)}$ initialized at $0$). The final layer's sequence $s_{1:T}^{(L)}$ carries the temporal context; a linear-sigmoid head
\begin{equation}
\hat{y}_t = \sigma\!\left(W_d\, s_t^{(L)} + b_d\right)
\label{eq:head}
\end{equation}
with head weight $W_d$ and bias $b_d$, maps it to the predicted probability $\hat y_t$ that the logical observable has flipped, evaluated at $t = T$ for the reported final-round prediction.

\subsection{Tile decomposition geometry}
\label{app:geometry}

Going from a $d=2k-1$ tile down to four $d=k$ tiles means scanning the four $k\times k$ corner regions of the larger tile's stabilizer lattice: each region reproduces exactly the layout of a standalone $d=k$ code, and stabilizers on a shared edge between two neighboring regions are included in both (diagonal regions share nothing), so the four tiles jointly cover every stabilizer of the parent code. Applying this once to a $d=9$ tile ($k=5$) gives four $d=5$ tiles; applying it again to each of those ($k=3$) gives four $d=3$ tiles per $d=5$ code, sixteen $d=3$ tiles in total.

\subsection{Tile aggregation}

Going up one code-distance level adds exactly one CNN. The base GNN encoder Eqs.~\eqref{eq:graphconv}--\eqref{eq:empty}, weights shared and continuing to train) runs independently on every one of $g\times g$ tiles covering the larger code, producing one embedding $h_c^{(a)} \in \mathbb{R}^{256}$ per window $c$ per tile $a = 1, \ldots, g^2$. These are arranged into a $g \times g \times 256$ spatial grid and combined by a two-layer CNN:
\begin{align}
O_c &= \mathrm{ReLU}\!\left( \sum_{a=1}^{g^2} W^{(a)} h_c^{(a)} + b_1 \right), \label{eq:conv1}\\
O_c' &= \mathrm{ReLU}\!\left(W_2\, O_c + b_2\right), \label{eq:conv2}
\end{align}
with weight matrices $W^{(a)}: \mathbb{R}^{256} \to \mathbb{R}^{H_{\rm agg}}$ (one per tile position $a$) and $W_2 \in \mathbb{R}^{H_{\rm agg} \times H_{\rm agg}}$, and biases $b_1, b_2$. The first layer's kernel size equals the tile-grid size, so it collapses the spatial dimensions to a single vector $O_c \in \mathbb{R}^{H_{\rm agg}}$ (equivalently, a per-tile linear map summed over tile positions), and the second is a $1\times1$ convolution, i.e.\ an ordinary linear layer. For $d = 9$, this repeats level by level: the lower level's aggregated embedding $O_c'$ is used in place of $h_c^{(a)}$ as the per-tile input to the next level's Eq.~\eqref{eq:conv1}, so each additional level adds one CNN Eqs.~\eqref{eq:conv1}--\eqref{eq:conv2} and nothing else. Recurrence over temporal windows Eq.~\eqref{eq:gru} is applied only once, at the outermost level, to the resulting top-level sequence $O_{1:T}'$ (hidden size $H_{\rm agg}$, $L_{\rm agg}$ layers); a linear-sigmoid head Eq.~\eqref{eq:head} on its final hidden state gives the logical-observable prediction.

\subsection{Training objective and optimization}

Every stage of Sec.~\ref{sec:pipeline} minimizes binary cross-entropy between the final-round prediction $\hat y_T$ Eq.~\eqref{eq:head} and the ground-truth logical-observable flip $y \in \{0, 1\}$,
\begin{equation}
\mathcal{L} = -\, y \log \hat y_T - (1 - y) \log(1 - \hat y_T),
\label{eq:loss}
\end{equation}
averaged over the batch and, for Stage 1 multi-$p$ training, over the sampled error rates. We minimize $\mathcal L$ with Adam~\cite{kingma2015adam}. Batch size defaults to $2048$ but is auto-tuned per run, searched over powers of two from $512$ to $16384$ to maximize GPU throughput, with the number of batches per epoch scaled inversely so the total stays fixed at $2048 \times 256 \approx 5.2 \times 10^5$ samples per epoch (default batch size times default batches per epoch); we use an exponentially decaying learning rate,
\begin{equation}
\mathrm{lr}(i) = \mathrm{lr}_0 \max\!\left(0.95^{\,i},\ 0.1\right),
\label{eq:lr_schedule}
\end{equation}
with $i$ the epoch index (reset between stages when fine-tuning from a checkpoint) and $\mathrm{lr}_0$ the stage's initial learning rate, floored at $10\%$ of $\mathrm{lr}_0$: $\mathrm{lr}_0 = 10^{-3}$ for Stage 1 synthetic pretraining, run for up to $600$ epochs, and $\mathrm{lr}_0 = 10^{-4}$ for Stage 2/3 fine-tuning, run for tens to a few hundred epochs, set per experiment by validation-accuracy plateau.

\section{Validation accuracy across the fine-tuning pipeline}
\label{app:val_from_start}
\begin{figure}
\includegraphics[width=\columnwidth]{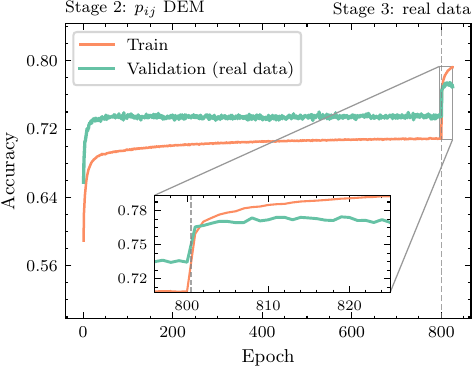}
\caption{Train accuracy (on the active stage's dataset) and real-data validation accuracy, evaluated from the start of Stage 2 through the end of Stage 3, for the $d=3$ memory patch q36 on \texttt{ibm\_berlin}. The dashed vertical line marks the Stage 2/3 transition; inset zooms into the last epochs, where validation accuracy plateaus even as train accuracy keeps rising.}
\label{fig:val_from_start}
\end{figure}
Fig.~\ref{fig:val_from_start} illustrates how much of the final real-data validation accuracy of the three-stage pipeline (Sec.~\ref{sec:pipeline}) is already reached during Stage 2 (synthetic fine-tuning on the calibrated detector error model) versus gained during Stage 3 (fine-tuning directly on real hardware detection events). For the $d=3$ memory patch (q36) on \texttt{ibm\_berlin}, we evaluate the held-out real-data validation split (Sec.~\ref{sec:pipeline}) at every epoch of both stages, starting from the same base model used throughout the paper: pretrained on Stim-simulated SI1000 noise (Stage 1). Validation accuracy rises from $65.8\%$ before fine-tuning to $73.5\%$ by the end of Stage 2 ($+7.7$ percentage points), and reaches $77.0\%$ by the end of Stage 3 ($+3.5$ additional percentage points): most of the accuracy gain over the Stage 1 pretrained base model comes from the synthetic calibrated-noise stage, with real hardware data contributing a smaller but consistent further improvement. Train accuracy, by contrast, keeps rising throughout Stage 3 (Fig.~\ref{fig:val_from_start}, inset) well past the point where validation accuracy plateaus, the usual signature of the model starting to overfit. This is exactly why validation accuracy, not train accuracy, selects the checkpoint used for all reported numbers: it tracks generalization to unseen real data without ever touching the held-out test set, so checkpoint selection cannot leak into the test accuracies reported elsewhere in this paper.

\bibliography{refs}

\end{document}